\documentclass[a4paper]{jpconf}

\usepackage{listings}

\usepackage{graphicx}
\usepackage[numbers,sort&compress,square]{natbib}
\usepackage{subcaption}
\usepackage[margin=2cm]{geometry}
\usepackage{hyperref}
\hypersetup{
    colorlinks=true,
    linkcolor=blue,
    filecolor=magenta,
    urlcolor=cyan,
    citecolor=blue,
}

\usepackage{xcolor}

\usepackage[acronym]{glossaries}
\glsdisablehyper
\usepackage{xspace}
\let\OldTexttrademark\texttrademark
\renewcommand{\texttrademark}{\OldTexttrademark\xspace}%
\usepackage{booktabs}
\usepackage{svg}

\usepackage{dsfont}
\usepackage{cancel}
\usepackage{bbm}

\usepackage{textcomp}

\newcommand{\px}[1]{\frac{\partial {#1}}{\partial x}}
\newcommand{\pz}[1]{\frac{\partial {#1}}{\partial z}}
\newcommand{\pxx}[1]{\frac{\partial^2 {#1}}{\partial x^2}}
\newcommand{\pzz}[1]{\frac{\partial^2 {#1}}{\partial z^2}}

\newacronym{les}{LES}{large eddy simulation}
\newacronym{scada}{SCADA}{supervisory control and data acquisition}
\newacronym{wat}{WAT}{wake-added turbulence}

\begin{document}

\title{Modelling Farm-to-Farm Interaction Using a Fast Linearised Numerical Approach}

\author{Alexia Everley$^{1,2}$, Hossein A. Kafiabad$^{2}$,  and Majid Bastankhah$^{1}$}

\address{$^1$Department of Engineering, Durham University, Durham, England\\
$^2$Department of Mathematical Sciences, Durham University, Durham, England}

\ead{alexia.everley@durham.ac.uk, hossein.amini-kafiabad@durham.ac.uk, majid.bastankhah@durham.ac.uk}


\vspace{1em}
\section*{Abstract}
This paper presents a computationally efficient, linearised numerical method for modelling aerodynamic interactions between wind farms. The linearised two-dimensional incompressible equations are solved using Fourier transforms in the horizontal direction and finite-difference discretisation in the vertical. Model predictions are validated against large-eddy simulation (LES) data, focusing on a tandem wind farm configuration where a downstream wind farm operates within the wake of an upstream array.
A parametric study is then conducted to examine the impact of this wake on the performance of the downstream farm across a range of inter-farm distances and hub-height ratios. We demonstrate that the upward vertical displacement of these wakes is driven by asymmetric turbulent entrainment caused by the farm's proximity to the ground, which restricts downward wake expansion. Consequently, the results suggest that, due to this upward wake displacement, downstream wind farms with higher hub heights may be more strongly affected by upstream farms than those with lower hub heights. 

\section{Introduction}
As the global deployment of offshore wind energy accelerates, the spatial footprint of wind power plants is rapidly expanding. This growth has led to increasing interest in the interactions between neighbouring wind farms, often referred to as farm-to-farm interactions \cite{bastankhah2024farmwake}. When multiple large farms are deployed in close proximity, the wake generated by an upstream farm can extend tens of kilometres downstream, reducing the available wind resource and thereby impacting the power production of downstream farms \cite{Nygaard2020, Pettas2021_interFarmInteractions}. As a result, the accurate modelling and prediction of farm-to-farm interactions has become a rapidly growing topic of interest within the wind energy community.

Modelling of wind farm flows has been typically pursued using two main approaches: high-fidelity computational fluid dynamics (CFD) simulations, most notably large-eddy simulations (LES), and engineering or analytical models \cite{stevens2017}. LES approaches resolve the unsteady three-dimensional turbulence dynamics governing wind farm wakes and can capture the detailed physical processes underlying wake recovery and interactions. However, they are computationally expensive, requiring substantial computational resources and long wall-clock times, which makes them less practical for large parametric studies, layout optimisation, or real-time applications. These computational demands become even more prohibitive when interactions among multiple wind farms are considered, as such simulations require computational domains that are significantly larger than those used for intra-farm simulations. On the other hand, engineering models are far more computationally efficient and can be run at minimal cost \cite{PorteAgel2020_review}. These models usually combine multiple components such as single-turbine wake models (e.g. Jensen \cite{jensen83}, Gaussian wake model \cite{bastankhah2014new} and their variants), wake superposition methods \cite{bastankhah2021analytical}, and blockage corrections \cite{bleeg2018blockage}. Coupling wake and blockage modelling is usually done in an iterative manner \cite{nygaard2020modelling} to estimate the farm-level flow field. While this approach is relatively inexpensive, the multi-component structure of such models can reduce their tractability. Moreover, they are often developed based on several simplifications, and the validity of these assumptions at large spatial scales, where interactions between wind farms and the atmosphere become significant, needs to be carefully evaluated.

This context motivates the development of an intermediate class of modelling tools that can bridge the gap between high-fidelity simulations and traditional engineering approaches by striking a more favourable balance between physical realism and computational efficiency. Fast, linearised numerical models are particularly attractive in this regard, as they can capture large-scale flow interactions at the wind-farm level while remaining computationally efficient. Such approaches have been explored in the community (\cite{ott2011Fuga,ott2014developments,ebenhoch2017}),  but their application has largely focused on intra-farm flow dynamics and wake interactions within individual wind farms. The work presented in this paper provides a preliminary framework which extends this class of models to larger-scale farm–atmosphere interactions. Specifically, we introduce a two-dimensional, steady-state formulation for a semi-infinite wind farm of finite length and infinite width, based on a set of systematic simplifying assumptions for turbine forcing and atmospheric turbulence. This allows us in particular to explore farm-farm interactions in a simple two wind farm tandem scenario.

The remainder of this paper is organised as follows. Section \ref{sec:math_form} presents the mathematical formulation of the proposed model and the governing equations underlying the linearised framework. Section \ref{sec:numerical_approach} describes the numerical approach adopted to solve the resulting system efficiently. The model performance and key physical insights are discussed in Section \ref{sec:results}. Finally, Section \ref{sec:conclusion} summarises the main conclusions of the study and outlines directions for future work.


\section{Mathematical formulation} \label{sec:math_form}
We consider the linearised Navier-Stokes equations around a shear flow base state just as in the derivation of the Orr-Sommerfeld equation \cite{Drazin_Reid_2004}.
Specifically, we consider the time-independent linearised two-dimensional incompressible equations following the seminal work of \cite{belcher-Hunt2003} on canopy flows,
\begin{subequations}\label{BLeq_linear}
\begin{align}
    U \px{u} + w \frac{d U}{d z} &= - \px{p} + \pz{\tau} + f, \label{x-momentum} \\
    U \px{w} &= - \pz{p}, \label{z-momentum} \\
    \px{u} + \pz{w} &= 0. \label{continuity}
\end{align}
\end{subequations}
Here, $u$ and $w$ are the horizontal and vertical velocity perturbations respectively and $p$ denotes the pressure perturbation divided by the reference density. $U = U(z)$ is the background velocity, which is assumed to have a logarithmic profile and asymptote to the wind speed aloft $U_{G}$. Only the vertical Reynolds shear stress perturbation is retained and denoted by $\tau$ (with the Reynolds normal stress gradients assumed negligible). For turbulence closure, we use $ \tau = \nu_t \, \partial {u}/ \partial z$, where the turbulent eddy viscosity coefficient, $\nu_t$, is taken to be constant for simplicity. 
However, our approach can be readily extended to spatially varying $\nu_t$.
The wind farm forcing (per unit mass) is denoted by $f$ in \eqref{x-momentum}. Since our focus in this study is not on resolving the detailed flow distribution within the wind farm but rather on capturing the large-scale flow behaviour downstream of it, we begin by assuming a simplified representation of the wind farm’s forcing similar to \cite{belcher-Hunt2003,smith_gravity_2010}. Instead of accounting for the contribution of each individual turbine, $f$ is assumed to be uniformly distributed across the entire farm area. Suppose the farm is arranged with streamwise inter-turbine spacing $S_x$ and spanwise inter-turbine spacing $S_y$, and each turbine has a rotor diameter $D$ (and radius of $R=D/2$) and a hub height $z_h$. Under these assumptions, $f$ can be expressed as
{\small
\begin{equation}\label{eq:farm_forcing}
    f=-\frac{N}{(N-1)}\frac{\pi D C_T \eta_w^2 U_h^2}{8 S_y S_x} \left[\mathcal{H}\left(z-(z_h-R)\right)-\mathcal{H}(z-(z_h+R))\right]\left[\mathcal{H}(x-x_0)-\mathcal{H}(x-(x_0+L_f))\right],
\end{equation}}
where $U_h=U(z=z_h)$ is the incoming velocity at the turbine hub height and $C_T$ is the turbine thrust coefficient which is assumed to be constant. $N$ is the number of turbine rows in the wind farm.
In reality, due to inflow shear and flow deceleration within the wind farm, the local incoming velocity experienced by wind turbines differs from $U_h$. To account for this effect while keeping the model simple, a constant parameter $\eta_w$ is introduced in \eqref{eq:farm_forcing} which, in practice, depends on the wind farm size, layout and turbine characteristic. As the main focus of this study is to develop the numerical framework, we treat the value of $\eta_w$ as a tuning parameter. In equation \eqref{eq:farm_forcing}, $\mathcal{H}$ denotes the Heaviside function, and $x_0$ represents the streamwise position of the leading edge of the farm (i.e. the location of the first turbine row). The farm length, denoted by $L_f$, is defined as $L_f = S_x (N-1)$. For numerical implementation, the Heaviside functions in equation \eqref{eq:farm_forcing} are replaced by ``$\tanh$" functions to ensure smoother variations in the applied forcing and to mitigate numerical instabilities. 

Differentiating \eqref{x-momentum} with respect to $z$ and \eqref{z-momentum} with respect to $x$ yields 
\begin{subequations}
    \begin{align}
        U \frac{\partial^2 u}{\partial z \partial x}+\frac{d U}{dz} \frac{\partial u}{\partial x}+ w \frac{d^2 U}{d z^2} + \frac{\partial w}{\partial z} \frac{d U}{d z} &= - \frac{\partial^2 p}{\partial z \partial x} + \pzz{\tau} + \frac{\partial f}{\partial z }, \label{x-momentum_difz} \\
    U \pxx{w} &= - \frac{\partial^2 p}{\partial x \partial z}. \label{z-momentum_difx}
    \end{align}
\end{subequations}
Using the continuity equation \eqref{continuity}, we have 
\begin{equation}\label{eq: integral cont}
    u = -\int_{-\infty}^x \pz{w}\,ds,
\end{equation}
assuming that far from the wind farm, as $x \rightarrow -\infty$, $u\rightarrow 0$. This leads to
\begin{equation}\label{taudef}
    \tau = -\nu_{t}\pz{} \bigg(\int_{-\infty}^x \pz{w}\,ds\bigg). 
\end{equation}
Subtracting \eqref{x-momentum_difz} from \eqref{z-momentum_difx} and using \eqref{continuity} and \eqref{taudef}, we obtain a partial differential equation (PDE) for $w$:
\begin{equation}\label{realBLeq4w}
U \left(\pzz{w} + \pxx{w} \right) - \frac{d^2 U}{d z^2} w - \left( \nu_t \int_{-\infty}^x \frac{\partial^4 w(z, s)}{\partial z^4} \, ds \right)= - \pz{f}.
\end{equation}

Periodic boundary conditions are imposed in the horizontal direction, since far from the farms $w$ and its derivatives vanish. In the vertical direction, the boundary conditions are specified as
\begin{equation}\label{eq:BC_physical}
\text{i)} \left. w \right|_{z=z_0} = 0, \quad \text{ii)} \left. \pz{w}\right|_{z=z_0} = 0, \quad \text{iii)} \left. \pz{w} \right|_{z=H} = 0 , \quad \text{iv)} \left. \pzz{w} \right|_{z=H} = 0,
\end{equation}
where $z=H$ marks the top of the boundary layer, at which the vertical variation of $w$ asymptotes to zero, resulting in the application of conditions iii) and iv) . Conditions i) and ii) follow directly from \eqref{continuity} together with the no slip boundary condition $u(x,z=z_0) = 0$, $w(x,z=z_0)=0$, where $z_0$ is the aerodynamic surface roughness.

\section{Numerical approach}\label{sec:numerical_approach}
To solve the PDE \eqref{realBLeq4w} developed in Section \ref{sec:math_form}, we use a spectral method in the horizontal direction, similar to implementations in prior studies \cite{ott2011Fuga}. In the vertical direction, we discretise the governing PDEs using a finite-difference method, allowing the model to resolve vertical flow variations that are averaged out in depth-integrated approaches (e.g., \cite{smith_gravity_2010, allaerts2019sensitivity}).

For the spectral method we utilise the following definitions of the discrete Fourier transform and its inverse
\begin{equation} \label{FT_definitions}
    \hat{q}_{k}(z) = \sum_{n=0}^{N_x} q(x_{n},z)e^{-ikx_{n}} \quad \text{and} \quad q(x_{n},z) = \frac{1}{N_{x}}\sum_{k=0}^{N_x} \hat{q}_{k}(z)e^{ikx_{n}},
\end{equation}
where $x_{n} = nD_x/N_{x}$ and $k = 2\pi k'/D_{x}$ with $D_{x}$ being the horizontal length of the domain, $N_{x}$ the number of Fourier modes and $k'$ the frequency index.
Taking the $x$-derivative of \eqref{realBLeq4w}, applying the (discrete) Fourier transform \eqref{FT_definitions} and then dividing by $ik$ for $k\neq0$ yields
\begin{equation}
    U(-k^2  \hat{w}_{k}+\frac{\partial^2  \hat{w}_{k}}{\partial z^2}) -  \hat{w}_{k} \frac{d^2U}{dz^2} +\frac{i\nu_{t}}{k}\frac{\partial^4\hat{w}_{k}}{\partial z^4} =  - \frac{\partial \hat{f}_k}{\partial z} \quad (\text{for} \quad k \ne 0). \label{spectralBLeq4w}
\end{equation}
$\hat{w}_{k}(z)$ and $\hat{f}_k(z)$ are the Fourier coefficients of $w(x,z)$ and $f(x,z)$ respectively at a given height $z$. To find an expression for the $k=0$ mode, we first consider the Fourier transform of \eqref{continuity} 
\begin{equation}
    ik\hat{u}_{k} + \pz{\hat{w}_{k}} = 0. \label{eq: FT cont}
\end{equation}
At $k=0$, this equation reduces to
\begin{equation}
     \pz{\hat{w}_{0}} = 0.
\end{equation}
Since on the lower boundary $w(x, z=z_0) = 0$, we conclude that $\hat{w}_0({z=z_0})=0$, leading to $\hat{w}_0(z) = 0$. 
We use $J$ points to discretise $\hat{w}_{k}$ and denote the discretised velocity field with $\hat{w}^i_k$ where $ i = 0,1, ..., J$. The vertical boundary conditions are implemented using a ghost point approach:
\begin{equation}
\text{i)}  \hat{w}_{k}^{-1}  = 0, \quad \text{ii)}  \frac{d\hat{w}_{k}^{-1}}{dz}= 0, \quad \text{iii)} \frac{d\hat{w}_{k}^{J+1}}{dz} = 0, \quad \text{iv)} \frac{d^2\hat{w}_{k}^{J+1}}{dz^2} = 0,
\end{equation}
where the indices $-1$ and $J+1$ denote the ghost points. The distance between consecutive grid points in the $z$-direction is represented by $\Delta z$. Applying the central difference approximations to the boundary conditions, we obtain
\begin{subequations}\label{discrestisedBC}
    \begin{align}
        &\text{i)}  \hat{w}_{k}^{-1}  = 0, \\ &\text{ii)}  \frac{d\hat{w}_{k}^{-1}}{dz}= \frac{\hat{w}_{k}^{0}- \hat{w}_{k}^{-2}}{\Delta z} = 0 \implies \hat{w}_{k}^{-2} = \hat{w}_{k}^{0}, \\&\text{iii)} \frac{d\hat{w}_{k}^{J+1}}{dz} = \frac{\hat{w}_{k}^{J+2}- \hat{w}_{k}^{J}}{\Delta z} = 0 \implies \hat{w}_{k}^{J+2} = \hat{w}_{k}^{J}, \\&\text{iv)} \frac{d^2\hat{w}_{k}^{J+1}}{dz^2} = \frac{\hat{w}_{k}^{J+2}-2\hat{w}_{k}^{J} + \hat{w}_{k}^{J+1}}{(\Delta z)^2} = 0 \implies \hat{w}_{k}^{J}-2\hat{w}_{k}^{J} + \hat{w}_{k}^{J+1} = 0\implies \hat{w}_{k}^{J+1} = \hat{w}_{k}^{J}.
    \end{align}
\end{subequations}
Imposing these boundary conditions, we can apply finite difference to find the discretised form of each term in \eqref{spectralBLeq4w}. 
Let $\mathcal{L}_1$, $\mathcal{L}_2$  $\mathcal{L}_3$  and $\mathcal{L}_4$ represent the discretised form of the operators $-k^2U$, $U \frac{\partial^2  \hat{w}_{k}}{\partial z^2} $, $-U''$ and $\frac{i\nu_{t}}{k}\frac{\partial^4\hat{w}_{k}}{\partial z^4}$ respectively. This results in the linear system 
\begin{equation}
    (\mathcal{L}_1 + \mathcal{L}_2+ \mathcal{L}_3+\mathcal{L}_4)[\hat{w}^i_k] = - \left[\frac{\partial \hat{f}_k^i}{\partial z}\right], \qquad i = 0,1, ..., J,
\end{equation}
which is solved efficiently via band-limited matrix inversion. The linear operators in matrix form are
\begin{gather*}
    \mathcal{L}_1= k^2\begin{bmatrix}
    U^{0}  & 0 & \\ 0 & U^{1}  & 0 &  \\ & \ddots & \ddots& \ddots \\
    & & 0 & U^{J-1} &0 \\ & && 0 & U^{J} 
\end{bmatrix}, \quad \mathcal{L}_2 = \frac{1}{(\Delta z)^2}\begin{bmatrix}
    -2U^{0} & U^{0} & 0 & \\ U^{1} & -2U^{1} & U^{1} & 0 & \\ 0 &U^{2} & -2U^{2} & U^{2} & 0 &  \\  &  & \ddots & \ddots& \ddots & \\
    & & 0 & U^{J-1} & -2U^{J-1} & U^{J-1} \\ & && 0 & U^{J} & -U^{J} 
\end{bmatrix} 
,\\
    \mathcal{L}_3 = \begin{bmatrix}
    U''^{0}  & 0 & \\ 0 & U''^{1}  & 0 &  \\ &\ddots & \ddots& \ddots \\
    & & 0 & U''^{J-1} &0 \\ & && 0 & U''^{J} 
\end{bmatrix} , \quad
\mathcal{L}_4= \frac{i\nu_{t}}{k(\Delta z)^4}\begin{bmatrix}
    7 & -4 & 1 & 0 &\dots \\ -4 & 6 & -4 & 1 & 0 & \dots \\ 1 & -4 & 6 & -4 & 1 & 0 & \dots \\ 0 & 1 & -4 & 6 & -4 & 1 & 0 & \dots \\ & \ddots & \ddots & \ddots & \ddots & \ddots & \ddots & \ddots \\ & &  0 & 1 & -4 & 6 & -4 & 1 \\ & & &  0 & 1 & -4 & 6 & -3 \\ &&& &  0 & 1 & -4 & 3  
\end{bmatrix} .
\end{gather*}
Note that we have employed the relations \eqref{discrestisedBC} in constructing the rows of $\mathcal{L}_2$ and $\mathcal{L}_4$ that are directly dependent on the boundary conditions. Once $\hat{w}_{k}$ has been computed for a finite set of wavenumbers, $w(x,z)$ is reconstructed using the inverse Fourier transform. The streamwise velocity field is subsequently derived by integrating \eqref{eq: integral cont} numerically.

\section{Results and Discussions}\label{sec:results}
\subsection{Model validation for a single wind farm}\label{sec:single_windfarm}
To build confidence in the proposed solver, it is necessary to both verify the validity of the linearised reduced formulation and the accuracy of the numerical solution.
For the former,  we compared predictions with the LES data presented in \cite{bastankhah2024farmwake}. Their simulations assume a semi-infinite wind farm that is infinite in the spanwise direction, making their laterally averaged streamwise velocity deficit suitable for comparison. We focus on their aligned baseline (B) and aligned short (S) cases, in which turbines are placed directly downstream of one another with fixed streamwise and spanwise spacing. The baseline case contains eight rows of turbines, whereas the short case contains four.
The parameters used for this comparison are shown in Table~\ref{tab:parameters MB}. $U_G$ is selected so that the hub height velocity is consistent with that used in the LES study. The value of normalised frictional velocity is stated as $u_*/U_h = 0.0296$. The turbulent viscosity coefficient is then estimated by applying the Prandtl mixing-length model to give $\nu_t = \kappa u_* z$. $\kappa$ is taken to be 0.4 and $z=z_h$ to obtain a constant value.

\begin{table}[t]
  \centering
  \caption{Summary of key parameters used in the single wind farm simulations of section \ref{sec:single_windfarm}.}
  \label{tab:parameters MB}
  \begin{tabular}{@{}lll@{}}
    \toprule
    Symbol & Description & Value \\
    \midrule
    $D$ & Turbine rotor diameter & $126\,\mathrm{m}$ \\
    $z_h$ & Turbine hub height & $90\,\mathrm{m}$ \\
    $C_T$ & Thrust coefficient & $0.776$ \\
    $\eta_w$ & Farm layout coefficient & B:$0.87$ \,\, S:$0.95$ \\
    $\nu_{t}$ & Turbulent viscosity coefficient & $8.5 \, \mathrm{m^2/s} $ \\
    $S_x$ & Streamwise spacing between turbines & $7D$ \\
    $S_y$ & Spanwise spacing between turbines & $4D$ \\
    $N$ & Number of turbine rows per farm & B:$8$ \,\, S:$4$  \\
    $L_f$ & Streamwise length of a single farm & B:$49D$ \,\, S:$21D$  \\
    $U_{G}$ & Atmospheric wind aloft & $9.45 \,\mathrm{m/s}$ \\
    $z_0$ & Surface roughness & $0.0002 \,\mathrm{m}$ \\
    $D_{x}$ & Domain length in $x$-direction &  $2000\,\mathrm{km}$\\
    $D_{z}$ & Domain length in $z$-direction &  $1\,\mathrm{km}$\\
    $N_{x}$ & Number of grid points in $x$-direction & $8192$ \\
    $N_{z}$ & Number of grid points in $z$-direction & $1000$\\
    \bottomrule
  \end{tabular}
\end{table}

\begin{figure}
     \centering
     \begin{subfigure}[b]{0.495\textwidth}
         \centering
         \includegraphics[width=\textwidth]{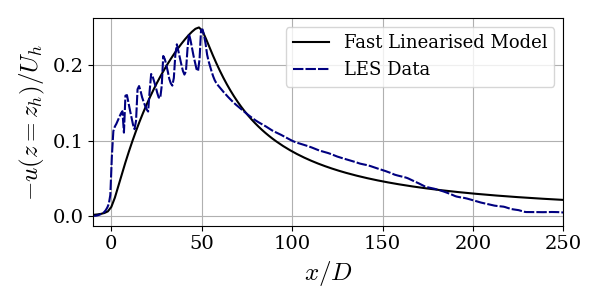}
         \caption{}
         \label{fig: Single Farm A0}
     \end{subfigure}
     \hfill
     \begin{subfigure}[b]{0.495\textwidth}
         \centering
         \includegraphics[width=\textwidth]{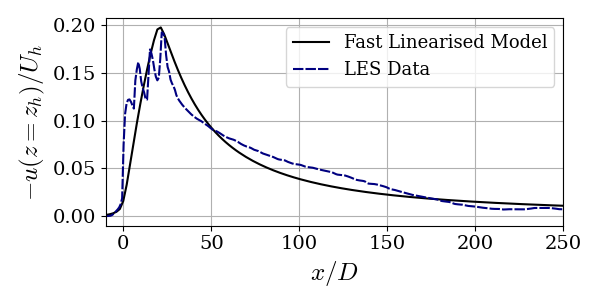}
         \caption{}
         \label{fig: Single Farm AS}
     \end{subfigure}
     \caption{Normalised streamwise velocity deficit at hub height for the (a) aligned baseline (B) and (b) aligned short (S) wind farm cases.}
     \label{fig: Single Farm}
\end{figure}

Figure~\ref{fig: Single Farm} illustrates the streamwise velocity deficit at hub height, normalised by $U_{h}$, for the two cases (B) and (S) in the left and right plots respectively. 
The figure shows that within the farm, the model is not able to capture velocity variations within each row, and especially the sudden increase in streamwise velocity deficit at the farm entrance where the first row of turbines is located. This is expected, as the simplified box forcing used in this study distributes the forcing across the entire farm footprint. The rate of change in deficit is captured by this fast linearised model better across the second half of each farm than the first.
Recovery of the wake begins immediately downstream of the farm in both simulations. The rate of this is slightly over-predicted by this model in both cases. These slight inaccuracies in results may be due to the simplifications made previously. 

\begin{figure}
     \centering
     \begin{subfigure}[b]{0.495\textwidth}
         \centering
         \includegraphics[width=\textwidth]{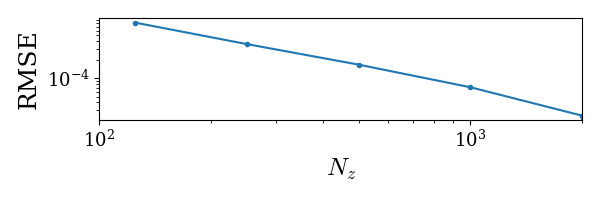}
         \caption{}
         \label{fig: RMS Nz varied}
     \end{subfigure}
     \hfill
     \begin{subfigure}[b]{0.495\textwidth}
         \centering
         \includegraphics[width=\textwidth]{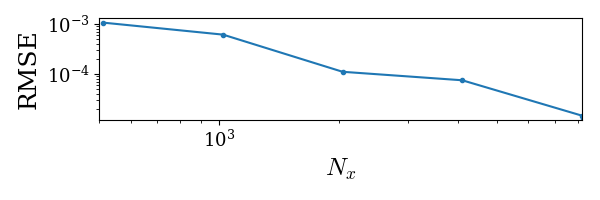}
         \caption{}
         \label{fig: RMS Nx varied}
     \end{subfigure}
     \caption{Root mean square error (RMSE) for a) $N_{x} = 16384$, $N_{z}$ varied and b)$N_{z} = 4000$, $N_{x}$ varied. True value is that taken for $N_{x} = 16384$ with $N_{z} = 4000$ and results are presented on a log-log scale.}
     \label{fig:RMS}
\end{figure}

To complete this two-part validation, the accuracy of the numerical solution can be verified by assessing its numerical convergence.
Figure~\ref{fig:RMS} shows the root-mean-square error (RMSE) obtained from this single farm scenario at different grid resolutions. The ``exact'' solution is taken as the result computed with the finest grid, $N_x = 16384$ and $N_z = 4000$.

Figure~\ref{fig: RMS Nz varied} shows the convergence with respect to the vertical resolution $N_z$. The RMSE decreases approximately linearly on the log-log scale as $N_z$ increases, dropping from $10^{-3}$ to $10^{-4}$ as $N_z$ increases from order $10^2$ to $10^3$. Figure~\ref{fig: RMS Nx varied} shows a similar behaviour for the streamwise resolution $N_x$, with the RMSE reaching values close to order $10^{-5}$ at the highest resolutions tested. 

These results indicate that resolutions of order $10^3$ in both directions provide a good balance between accuracy and computational cost. At these resolutions, the simulations remain extremely inexpensive: each case shown in Figure~\ref{fig: u for 3 cases} requires approximately $5$ seconds to run on a standard laptop (11th Gen Intel(R) Core(TM) i5-1135G7 @ 2.40GHz with 8~GB RAM). With confidence in the proposed solver established, we proceed to examine a tandem farm scenario.

\subsection{Tandem farm case}\label{sec:tandem}
To investigate the influence of upstream wind farms on the performance of downstream farms, we consider a configuration consisting of two identical wind farms arranged in tandem. The downstream farm is placed at varying distances from its upstream counterpart to examine how inter-farm spacing affects the wake interactions and overall performance. Following the LES investigation set-up in ~\cite{Stieren_Stevens_2022}, the spacing between the two farms, denoted by $L_x$, is defined as the streamwise distance between the trailing edge of the upstream farm and the leading edge of the downstream farm. We investigate three values of $L_x$: (i) $5\,\mathrm{km}$, (ii) $10\,\mathrm{km}$, and (iii) $15\,\mathrm{km}$. For each distance, a staggered wind farm configuration is considered, and for case (ii) an additional aligned configuration is also examined. This is accounted for by utilising a different $\eta_w$ for each configuration type.
We compare our 2D model predictions with the laterally averaged LES results at hub height. It is worth noting that, although the model developed here does not account for the lateral energy entrainment present in finite wind farms, for deep boundary layers such as those considered in~\cite{Stieren_Stevens_2022} the vertical entrainment is expected to dominate. Hence, the model should still provide reasonably realistic results.
The friction velocity $u_*$ for this study is determined via two methods. One estimates $u_*$ by fitting a log profile $U = ({u_*}/{\kappa})\ln{(z/z_0)}$ to the given inflow velocity profile and the other, by
using the reported wall shear stress $\tau_w$ where  $u_* \approx\sqrt{\tau_w/\rho}$; in this case $\rho$ has been separately accounted for. These approaches yield values of 
$\nu_t=14.1$ and $13.9$, respectively. Considering the close agreement between these values, we adopt $\nu_t = 14.0$ in our simulations. The geostrophic velocity $U_G$ is then chosen so that the hub-height velocity equals $9.5 \, \mathrm{m^2/s}$. More information on the numerical setup is provided in Table \ref{tab:parameters}. As shown in Table \ref{tab:parameters}, two distinct values of $\eta_w$ were used for the staggered and aligned wind farms. The domain specifications follow those listed in Table~\ref{tab:parameters MB}. 

\begin{table}[t]
  \centering
  \caption{Summary of key parameters used in the tandem wind farm simulations of section \ref{sec:tandem}.}
  \label{tab:parameters}
  \begin{tabular}{@{}lll@{}}
    \toprule
    Symbol & Description & Value \\
    \midrule
    $D$ & Turbine rotor diameter & $120\,\mathrm{m}$ \\
    $z_h$ & Turbine hub height & $100\,\mathrm{m}$ \\
    $C_T$ & Thrust coefficient & $0.75$ \\
    $\eta_w$ & Farm layout coefficient & Stag: $1.07$ Align: $0.95$ \\
    $S_x$ & Streamwise spacing between turbines & $7D$ \\
    $S_y$ & Spanwise spacing between turbines & $5D$ \\
    $\nu_{t}$ & Turbulent viscosity coefficient & $ 14.0 \, \mathrm{m^2/s} $\\
    $N$ & Number of turbine rows per farm & $12$ \\
    $L_f$ & Streamwise length of a single farm & $9.24 \,\mathrm{km}$ \\
    $L_x$ & Inter-farm spacing & $5 \,\mathrm{km},\; 10\,\mathrm{km},\; 15\,\mathrm{km}$ \\
    $U_{G}$ & Atmospheric wind aloft & $11.5 \,\mathrm{m/s}$ \\
    $z_{0}$ & Surface roughness & $0.002 \,\mathrm{m}$ \\
    \bottomrule
  \end{tabular}
\end{table}

\begin{figure}
    \centering
    \includegraphics[width=0.75\linewidth]{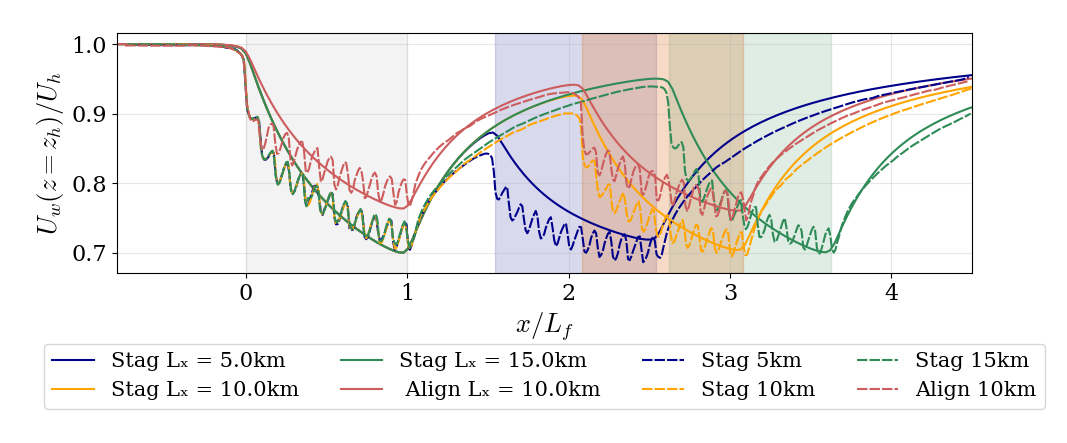}
    \caption{Normalised streamwise velocity at hub height for inter-farm distances $5 \mathrm{km}$, $10 \mathrm{km}$ and $15 \mathrm{km}$ (turbines staggered)  and $10 \mathrm{km}$ (turbines aligned). LES results are represented with a dashed line and our model's results with a solid line. The shaded regions indicate the location of each farm.}
    \label{fig: AS u_norm}
\end{figure}

In Figure~\ref{fig: AS u_norm}, we compare 
the hub-height value of the streamwise velocity $U_w=U+u$ normalised by $U_h$.
The figure shows that, as in the single farm case, the decrease in velocity is better captured across the second half of each farm than the first. The wake evolution downstream of each farm is generally well represented by our model. Both the velocity deficit and wake recovery resulting from the first farm is better captured than from the second. This discrepancy may be explained by our assumption of a constant turbulent eddy viscosity ($\nu_t$). In reality, turbulence changes considerably as the flow passes through the two farms, creating a complex spatial distribution of the turbulent viscosity that a constant value of turbulent viscosity may not realistically capture.

\begin{figure}
    \centering
    \includegraphics[width=0.6\linewidth]{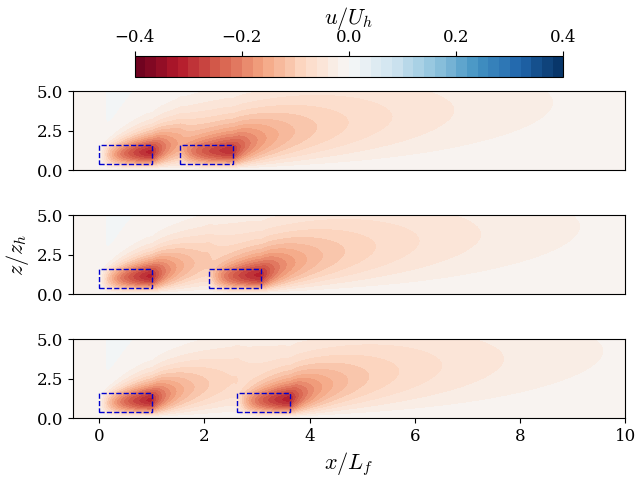}
    \caption{Contours of normalised streamwise velocity perturbation in the $xz$ plane for inter-farm distances of $5 \mathrm{km}$, $10 \mathrm{km}$ and $15 \mathrm{km}$. The dashed blue lines represent the location of the wind farms.}
    \label{fig: u for 3 cases}
\end{figure}

With the model predictions validated, we discuss the results in more detail below to provide some physical insights into the evolution of the farm wakes. 
Figure~\ref{fig: u for 3 cases} shows the vertical cross-sectional distribution of the streamwise velocity perturbation for the three different tandem staggered wind farm configurations. 
In all cases, a sharp increase in streamwise velocity deficit occurs at the position of the first farm, followed by a gradual recovery as turbulence mixes high-momentum air from the surroundings into the wake. 
As expected, the amount of wake recovery occurring prior to the entrance of the second farm increases with downstream spacing. When the second farm is placed only $5 \,\mathrm{km} \, (L_x \approx 0.5L_f)$ downstream, it is located mostly within the core of the upstream wake, where the velocity deficit remains large and the available energy is significantly reduced. For the case of $10 \,\mathrm{km} \, (L_x \approx L_f)$ inter-farm distance, the wake has partially recovered, resulting in a smaller velocity perturbation at the location of the second farm. At $15 \,\mathrm{km} \, (L_x \approx 1.6L_f)$ downstream spacing, the velocity deficit has recovered further, and the second farm experiences a much less disturbed inflow.

A notable feature of Figure~\ref{fig: u for 3 cases} is the apparent upward displacement of the wake centre. Although this behaviour has been reported in several recent LES studies \citep{wangUpwardShiftWind2023,kasper2024simulation,souaibyAtmosphericStabilityEffect2025}, there is still no clear consensus regarding the mechanism responsible for it. Leveraging the low computational cost of our linearised model, which allows controlled numerical experiments that isolate the influence of individual parameters, we investigate the origin of this upward tilting.

\begin{figure}[h!]
    \centering
    \begin{subfigure}[b]{0.62\textwidth}
        \centering
        \includegraphics[width=\textwidth]{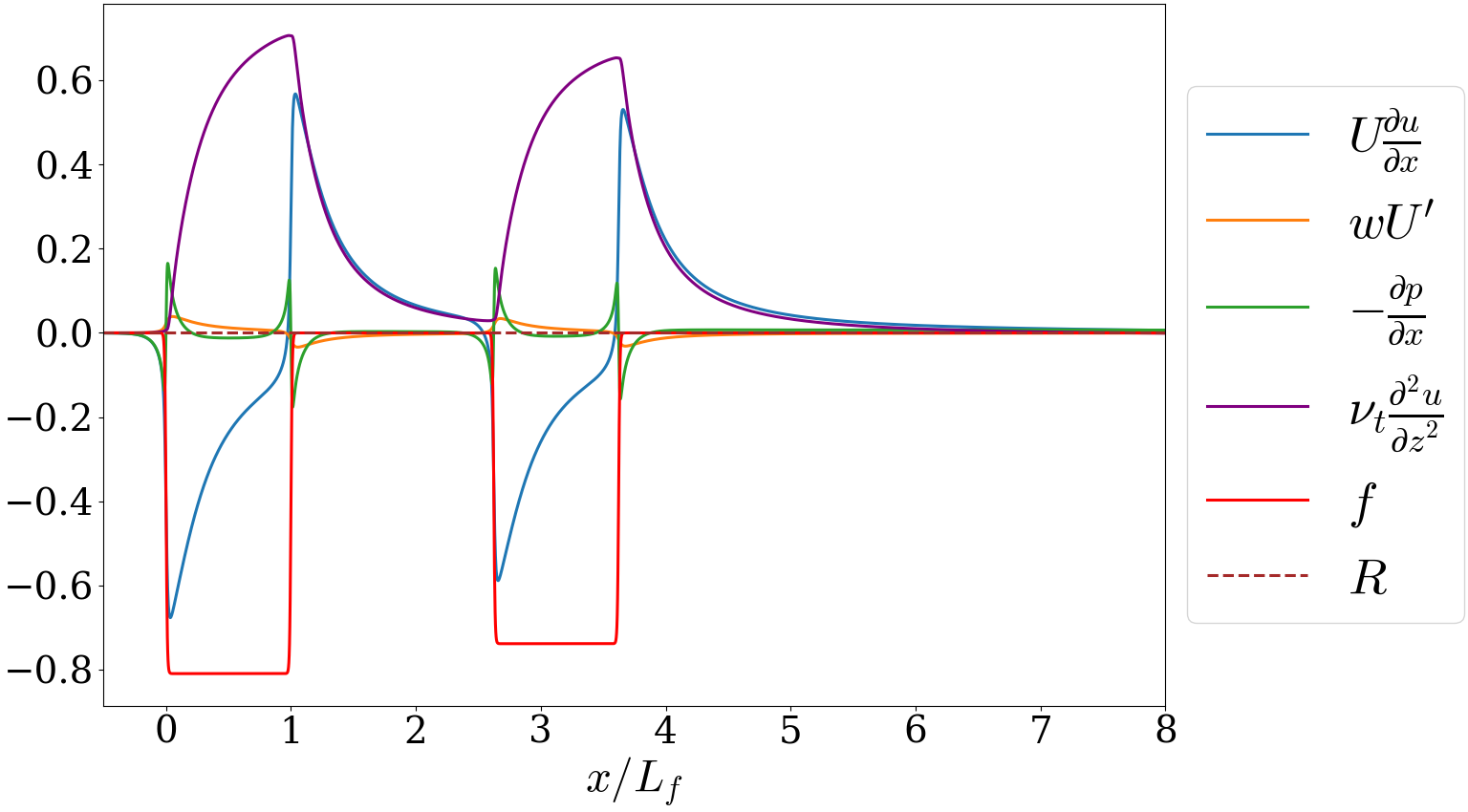}
        \caption{}
        \label{fig: Budget Analysis}
    \end{subfigure}
    \hfill 
    \begin{subfigure}[b]{0.36\textwidth}
        \centering
        \includegraphics[width=\textwidth]{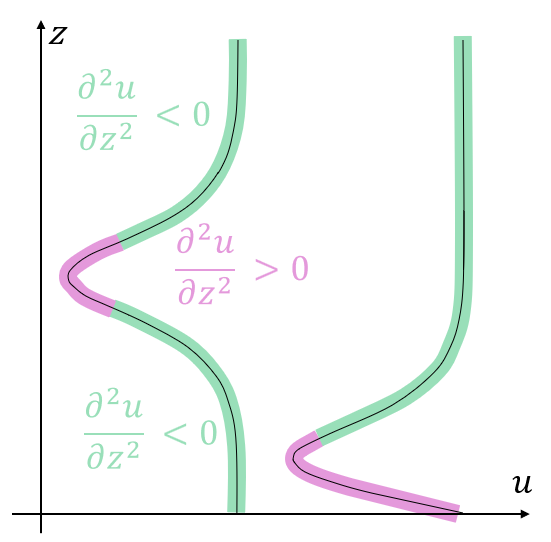}
        \caption{}
        \label{fig: Schematic for u}
    \end{subfigure}
    \caption{(a) Budget analysis conducted at $z=z_h$ for \eqref{x-momentum}, where the term $R$ is the residual. Each term is non-dimensionalised by $U_h^2/L_f$. (b) Schematic of $u$ profile at the end of the first farm. The left line represents the scenario for $z_h = D_z/2$ and the right line $z_h = D_z/10$.}
    \label{fig: budget_and_schematic}
\end{figure}

Kasper \textit{et al.}~\cite{kasper2024simulation} argued that the phenomenon is not caused by upward advection of the wind-farm wake. Instead, they attributed it to wake recovery driven by height-dependent vertical entrainment associated with the shear of the incoming boundary layer. Our results confirm their observation that the vertical displacement is primarily caused by asymmetric entrainment rather than by vertical advection. As shown in Figure~\ref{fig: Budget Analysis}, a budget analysis of equation~\eqref{x-momentum} indicates that, in the wake of each wind farm, the advection term governing the streamwise evolution of the velocity is predominantly balanced by turbulent entrainment,
\begin{equation}\label{eq:x_mom_farwake}
U\frac{\partial u}{\partial x} \approx \nu_{t}\frac{\partial^{2}u}{\partial z^{2}}.
\end{equation}

Consequently, since the streamwise recovery rate $({\partial u}/{\partial x})$ is controlled by turbulent entrainment, any vertical asymmetry in this term results in height-dependent wake recovery and therefore a displacement of the wake centre. The next question is what gives rise to this asymmetry. To address this, we exploit the ability of the linearised model to decouple the influence of background shear from other parameters. Additional simulations were therefore conducted using a uniform background flow instead of a boundary-layer profile. Even under uniform inflow conditions (results not shown here), the asymmetric wake recovery and vertical shift remained clearly visible. This allows us to rule out inflow shear as the primary driver of the wake displacement.

Instead, we postulate that the asymmetric entrainment arises from the proximity of the farm forcing to the ground. This mechanism can be understood geometrically by examining the curvature of the vertical profiles of the streamwise velocity perturbation, i.e.\ $\partial^{2}u/\partial z^{2}$. Figure~\ref{fig: Schematic for u} illustrates vertical profiles of $u$ (with curvature indicated by shading) for two scenarios: a farm located near the centre of the boundary layer and a farm located close to the ground. In the first case, the curvature is positive at the wake centre ($\partial^{2}u/\partial z^{2} > 0$) and negative at the upper and lower edges ($\partial^{2}u/\partial z^{2} < 0$). According to equation~\eqref{eq:x_mom_farwake}, the positive curvature at the wake centre accelerates the flow and promotes wake-centre recovery, while the negative curvature at the edges decelerates the flow at those locations and leads to wake expansion. 

When the wake centre is close to the ground, however, the available region beneath the hub height is strongly restricted. As a result, there is almost no region of negative curvature below the wake centre, leaving limited scope for flow deceleration and downward wake expansion. Consequently, as the wake evolves downstream, the upper part of the wake expands more rapidly than the lower part, producing the apparent upward shift seen in Figure~\ref{fig: u for 3 cases}. To verify this interpretation, additional simulations were performed with forcing placed at the centre of the boundary layer. In that configuration (results not shown here), the wake recovery remained symmetric, regardless of whether a free-surface or rigid-lid top boundary condition was applied. This suggests that the vertical position of the farm forcing relative to the boundary-layer thickness is the dominant cause of the asymmetric entrainment, rather than the shear in the inflow profile.

Although the above geometric explanation sheds light on the physical mechanism driving the wake's upward shift, the quantitative results should be interpreted with caution. The extent of asymmetric entrainment is expected to be sensitive to the turbulence closure model. Because this study assumes an overly simplified constant eddy viscosity $\nu_{t}$, further research using more advanced turbulence models is required to accurately quantify the magnitude of this displacement.

\begin{figure}
    \centering
    \includegraphics[width=0.6\linewidth]{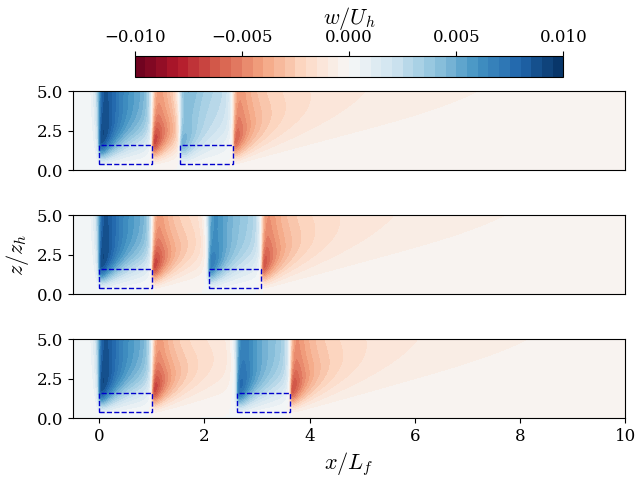}
    \caption{Contours of normalised vertical velocity perturbation in the $xz$ plane for inter-farm distances of $5 \mathrm{km}$, $10 \mathrm{km}$ and $15 \mathrm{km}$. The dashed blue lines represent the location of the wind farms.}
    \label{fig: w for 3 cases}
\end{figure}

Following the previous analysis focused on the streamwise velocity perturbation, examining the vertical velocity provides a complementary view of the dynamics. Figure~\ref{fig: w for 3 cases} shows the vertical velocity perturbation induced by the wind farms. Positive perturbations occur at and above the farms, indicating upward deflection of the inflow as it encounters the turbines, with the strongest effect at the upstream farm. Downstream of each farm, the perturbation becomes negative, corresponding to a downward motion of air. As the separation distance between the farms increases, the vertical acceleration at the second farm becomes stronger, likely due to increased wake recovery before the flow reaches the entrance of the downstream farm. The vertical velocity perturbations are also approximately one order of magnitude smaller than the corresponding streamwise perturbations.

\subsection{Parametric study for varied hub height of the downstream farm}
\begin{figure}
    \centering
    \includegraphics[width=0.6\linewidth]{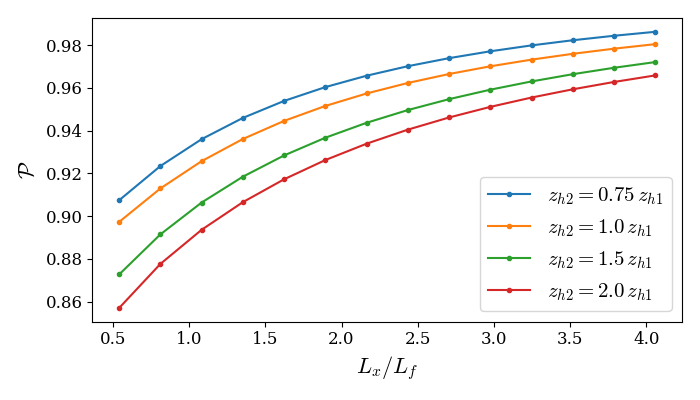}
    \captionsetup{skip=0.5pt}
    \caption{Ratio of power availability, $\mathcal{P}$, of a downstream wind farm operating in the wake of an upstream wind farm as a function of the normalized inter-farm spacing $L_x/L_f$, for different hub-height ratios $z_{h2}/z_{h1}$. $\mathcal{P}$ is normalized by the power of the downstream wind farm operating in isolation.
}
    \label{fig: Zh2 varied}
\end{figure}

To exploit the computational efficiency of the model developed here, we perform a parametric study to investigate how the potential power production of the downstream farm depends on the inter-farm separation distance and the ratio of turbine hub heights. Assuming that the available power scales with the cube of the local wind speed, $\mathcal{P}$ is defined as the ratio of the volume-averaged cube of the velocity within the downstream wind farm region when an upstream farm is present to that when the downstream farm operates in isolation (i.e. in the absence of any upstream farm). Figure~\ref{fig: Zh2 varied} presents $\mathcal{P}$ as a function of the inter-farm separation. The hub heights of the upstream and downstream wind farms are denoted by $z_{h1}$ and $z_{h2}$, respectively.

In all cases, increasing the separation distance leads to an increase in the potential power of the waked farm, bringing it closer to that of an isolated farm. This behaviour reflects the progressive recovery of the wind-farm wake discussed earlier. However, as shown in Section~\ref{sec:tandem}, the wake centre shifts upward as the wake spreads downstream. Consequently, the region most strongly affected by the upstream farm lies above the hub height of the upstream turbines. As a result, increasing the hub height of the downstream farm leads to a reduction in power generation. This observation suggests that newer wind farms, which are typically equipped with taller turbines, may be more strongly affected by older upstream farms than vice versa. As the present study assumes a simplified constant eddy viscosity $\nu_t$, the quantitative magnitude of this displacement should be interpreted with caution.

\section{Conclusion}\label{sec:conclusion}

We have presented a fast linearised numerical approach for modelling farm-to-farm interactions using a mixed spectral–finite-difference solution of the linearised two-dimensional incompressible equations. The model provides an intermediate framework between high-fidelity LES and traditional engineering wake models, retaining key physical mechanisms while remaining computationally inexpensive. Given its two-dimensional nature, the model is expected to perform best for deep boundary layers where vertical entrainment dominates and for very wide wind farms where edge effects do not significantly influence the main array. 

The method reproduces several features observed in LES studies of tandem wind farms. In particular, increasing the inter-farm spacing allows greater wake recovery and therefore reduces the impact on the downstream farm. Despite the simplicity of the uniform top-hat forcing representation, the model captures both the wind deceleration induced by each farm and the subsequent downstream wake recovery. Our analysis further shows that the apparent upward displacement of the farm wake arises from an imbalance in turbulent entrainment above and below the hub height. Because the farm forcing is located close to the ground, the region beneath the wake centre is restricted, limiting downward wake expansion and causing the upper portion of the wake to expand more rapidly. This behaviour can be interpreted geometrically through the curvature of the vertical velocity profiles.

Exploiting the computational efficiency of the linearised model, we also performed a parametric study of the potential power ratio between waked and isolated wind farms for different hub-height combinations. Because the wake centre of the upstream farm shifts upward as it propagates downstream, increasing the hub height of the downstream farm reduces the potential power ratio. This suggests that newer wind farms equipped with taller turbines may experience stronger wake impacts from older upstream farms, although further investigation is required to quantify this effect.

The framework presented also resolves vertical velocity perturbations within the atmospheric boundary layer, enabling more realistic investigation of processes involving vertical motion, such as gravity waves, beyond the limitations of depth-averaged models. A key advantage of the approach is its computational efficiency: accurate solutions can be obtained using relatively coarse grids with minimal cost, allowing simulations to be completed within seconds on a standard laptop. This makes the method well suited for applications requiring rapid evaluation, such as parametric studies or wind-farm layout design. 

We view the present work as a proof of concept with strong potential for further development. Immediate directions include extending the framework to incorporate three-dimensional effects and improving the representation of parameters such as turbulent viscosity and turbine forcing. 

\bibliographystyle{iopart-num}
\bibliography{references}

\end{document}